\documentclass[twocolumn,showpacs,preprintnumbers]{revtex4}
\usepackage{graphicx}
\usepackage{dcolumn}
\usepackage{bm}

\input{tcilatex}

\begin{document}

\title{Exact uncertainty principle and quantization: implications for the
gravitational field}
\author{M. Reginatto}
\email{Marcel.Reginatto@ptb.de}
\affiliation{Physikalisch-Technische Bundesanstalt, Bundesallee 100, 38116 Braunschweig,
Germany}
\date{\today}

\begin{abstract}
The quantization of the gravitational field is discussed within the exact
uncertainty approach. The method may be described as a Hamilton-Jacobi
quantization of gravity. It differs from previous approaches that take the
classical Hamilton-Jacobi equation as their starting point in that it
incorporates some new elements, in particular the use of a formalism of
ensembles in configuration space and the postulate of an exact uncertainty
relation. These provide the fundamental elements needed for the transition
from the classical theory to the quantum theory.
\end{abstract}

\maketitle


%



\section{Introduction}

Perhaps one of the most fundamental distinctions between classical and
quantum systems is that a quantum system must obey the uncertainty
principle, while a classical system is not subject to such a limitation. The
degree to which a quantum system is subject to this limitation is given by
Heisenberg's uncertainty relation, which provides a bound on the minimum
uncertainty of two conjugate variables such as the position\ $x$ and the
momentum $p$ of a particle: it states that the product of their
uncertainties must be equal to or exceed a fundamental limit proportional to
Planck's constant, $\Delta x\Delta p\geq \hbar /2$. This inequality,
however, is not sufficiently restrictive to provide a means of going from
classical mechanics to quantum mechanics. Indeed, in most approaches to
quantization, no direct reference is made to the \textit{need} for an
uncertainty principle. For example, in the standard approach to canonical
quantization, one introduces a Hilbert space, operators to represent
observables, and a representation of the Poisson brackets of the fundamental
conjugate variables in terms of commutators of their corresponding
operators. Once this structure is in place, one can show that the Heisenberg
uncertainty relations are indeed satisfied -- but the uncertainty relations
are \textit{derived}, and the uncertainty principle, instead of motivating
the method of quantization, seems to be little more than a consequence of
the mathematical formalism.

One can, however, introduce an \textit{exact} form of the uncertainty
principle which is strong enough to provide a means of going from classical
to quantum mechanics -- indeed, this exact uncertainty principle provides
the \textit{key} element needed for such a transition. In particular, the
assumption that a classical ensemble of particles is subject to nonclassical
momentum fluctuations, of a strength inversely proportional to uncertainty
in position, leads directly from the classical equations of motion to the
Schr\"{o}dinger equation \cite{HR02A,HR02B}. This approach can also be
generalized and used to derive bosonic field equations \cite{HKR03}. The
exact uncertainty approach is extremely minimalist in nature: unlike
canonical quantization, no \textit{a priori} assumptions regarding the
existence of a Hilbert space structure, linear operators, wavefunctions,
etc. is required. The sole \textquotedblleft nonclassical\textquotedblright\
element needed is the addition of fluctuations to the momentum variable. The
exact uncertainty approach is thus rather economical, in that it appears to
make use of the minimum that is required for the description of a quantum
system.

The paper is organized as follows. In the next two sections, I briefly
review the case of a non-relativistic particle: in section 2,\ I discuss
classical ensembles and derive the equations of motion from an ensemble
Hamiltonian; in section 3, I present the derivation of the Schr\"{o}dinger
equation using the exact uncertainty approach. In the remaining sections, I
discuss the application of the exact uncertainty approach to gravity. A
recent paper \cite{H04} provides an overview of the exact uncertainty
approach in quantum mechanics and quantum gravity that emphasizes conceptual
foundations. Therefore, in the last sections I focus instead on some
technical issues and on some issues of interpretation. In section 4, I
introduce the Hamilton-Jacobi formulation of general relativity and
formulate a theory of classical ensembles of gravitational fields; in
section 5, I apply the exact uncertainty principle to the gravitational
field, which leads to a Wheeler-DeWitt equation. In section 6, I consider
some implications of the exact uncertainty approach for the description of
space-time in the quantized theory. Finally, some further remarks concerning
the application of the exact uncertainty approach to gravity are given in
section 7.


\section{classical ensembles of non-relativistic particles}

To motivate the introduction of classical ensembles in configuration space,
consider the possibility that, due to either practical or theoretical
reasons, the position of a particle is not known exactly. To describe this
uncertainty in position, one has to introduce a probability density $P\left(
x,t\right) $ on configuration space. The description of the physical system
is then one that is given in terms of a \textit{statistical ensemble} of
particles.\ The dynamics of such an ensemble may be described using a
Hamiltonian formalism, in the following way. For an ensemble of classical,
non-relativistic particles of mass $m$ moving in a potential $V$, the
correct equations of motion can be derived from the ensemble Hamiltonian 
\begin{equation}
\tilde{H}_{c}\left[ P,S\right] =\int dxP\left( \frac{\left\vert \nabla
S\right\vert ^{2}}{2m}+V\right) .  \label{Eq01}
\end{equation}%
The equations of motion that follow from eq. (\ref{Eq01}),%
\[
\frac{\partial P}{\partial t}=\frac{\delta \tilde{H}_{c}}{\delta S},\quad 
\frac{\partial S}{\partial t}=-\frac{\delta \tilde{H}_{c}}{\delta P}, 
\]%
take the form%
\begin{equation}
\frac{\partial P}{\partial t}+\nabla \cdot \left( P\frac{\nabla S}{m}\right)
=0,  \label{Eq02}
\end{equation}%
\begin{equation}
\frac{\partial S}{\partial t}+\frac{\left\vert \nabla S\right\vert ^{2}}{2m}%
+V=0.  \label{Eq03}
\end{equation}%
Eq. (\ref{Eq02}) is a continuity equation\ for an ensemble of particles
described by the velocity field $v=m^{-1}\nabla S$, and eq. (\ref{Eq03}) is
the Hamilton-Jacobi equation for a classical particle with momentum $%
p=\nabla S$. Note also that $\tilde{H}_{C}$ is the average energy of the
classical ensemble. The formalism of ensemble Hamiltonians that I have used
here, while very simple, turns out to be quite general (see \cite{H04A,H04}
for more details on this formalism and various applications).\ In
particular, it can be shown that both classical and quantum equations of
motion can be derived using this formalism. The difference between classical
and quantum ensembles can be characterized by a simple difference in the
functional form of the corresponding ensemble Hamiltonians, as shown in the
next section.


\section{quantum equations for non-relativistic particles}

To derive quantum equations of motion, one needs to add fluctuations to the
momentum variable which satisfy an exact uncertainty principle, and modify
the classical ensemble Hamiltonian to take into consideration the effect of
these fluctuations. There are, of course, further assumptions that need to
be made, and these are explained in this section -- however, the main
conceptual ingredients are the nonclassical momentum fluctuations and the
exact uncertainty principle.

Consider then another type of ensemble, one for which the assumption of a
deterministic relation between position and momentum (implicit in the
classical relation $p=\nabla S$) is no longer valid. Assume instead that the
momentum is subject to stochastic perturbations, and that the relation
between $p$ and $\nabla S$ takes the form%
\[
p=\nabla S+f 
\]%
where the fluctuation field $f$ vanishes everywhere on average. Denoting the
average over fluctuations by an overline, and using the relations $\overline{%
f}=0$ and $\overline{p}=\nabla S$, the classical ensemble energy is replaced
in this case by%
\begin{eqnarray*}
\left\langle E\right\rangle &=&\int dxP\left( \frac{\overline{\left\vert
\nabla S+f\right\vert ^{2}}}{2m}+V\right) \\
&=&\tilde{H}_{c}+\int dxP\frac{\overline{f\cdot f}}{2m}
\end{eqnarray*}%
Thus the momentum fluctuations contribute to the average kinetic energy of
this modified classical ensemble.

What conditions on the fluctuation field $f$ lead to the quantum equations
of motion? It is sufficient to consider four principles \cite%
{HR02A,HR02B,HKR03} (i) \textit{Action principle}: The equations of motion
can be derived from an action principle based on an ensemble Hamiltonian.
Then, the strength of the fluctuations is determined by some function of $P$
and $S$ and their derivatives, i.e. $\overline{f\cdot f}=\alpha \left(
x,P,S,\nabla P,\nabla S,...\right) $. It will also be assumed that the
equations of motion that result are partial differential equations of at
most second order. (ii) \textit{Independence}: if the system consists of two
independent non-interacting sub-systems, the ensemble Hamiltonian, and
therefore $\alpha $, decomposes into additive subsystem contributions. (iii) 
\textit{Invariance}: the fluctuations transform correctly under linear
canonical transformations (i.e., $f\rightarrow L^{T}f$ for any invertible
linear coordinate transformation $x\rightarrow Lx$. A sightly different
principle was used in \cite{HR02A,HR02B}). (iv) \textit{Exact uncertainty
principle}: the strength of the momentum fluctuations is determined solely
by the uncertainty in position. Since $P$ contains all the information on
the position uncertainty, $\alpha $ can only depend on $P$ and its
derivatives.

The first three principles are natural on physical grounds, and therefore
relatively unconstraining. The fourth principle, however, by establishing a
precise relationship between the position uncertainty (as described by $P$)
and the strength of the momentum fluctuations, introduces a strongly \textit{%
nonclassical} element. It is therefore appropriate to refer to the
fluctuations due to the field $f$ as nonclassical\ momentum fluctuations. It
can be shown that these four principles define the functional form of $%
\alpha $ uniquely and lead to a quantum ensemble Hamiltonian of the form%
\begin{equation}
\tilde{H}_{q}=\tilde{H}_{c}+C\int dx\frac{1}{P}\frac{\left\vert \nabla
P\right\vert ^{2}}{2m}  \label{Eq04}
\end{equation}%
where $C$ is a positive universal constant. Furthermore, since the term $%
\alpha $ is proportional to the classical Fisher information, one can use
the Cram\'{e}r-Rao inequality of statistical estimation theory to derive an
exact uncertainty relation that is stronger than (and implies) the
Heisenberg uncertainty relation (see \cite{HR02A,HR02B,HKR03} for
derivations of these results).

It is straightforward to show that the Hamiltonian equations of motion for $%
P $ and $S$ derived from eq. (\ref{Eq04}),%
\[
\frac{\partial P}{\partial t}=\frac{\delta \tilde{H}_{q}}{\delta S},\quad 
\frac{\partial S}{\partial t}=-\frac{\delta \tilde{H}_{q}}{\delta P}, 
\]%
are identical to the Schr\"{o}dinger equation for a wavefunction $\psi $,
provided one defines%
\[
\hbar :=2\sqrt{C},\quad \psi :=\sqrt{P}e^{iS/\hbar }. 
\]%
As shown in \cite{HR02A}, the wavefunction representations arises naturally
from seeking canonical transformations which map the fields $P$ and $S$ to
the \textquotedblleft normal modes\textquotedblright\ of the system. This
observation, together with a symmetry assumption, leads then to the usual
Hilbert space formulation of quantum mechanics.


\section{classical ensembles of gravitational fields}

The exact uncertainty approach can be succesfully generalized to derive the
equations of motion for bosonic fields with Hamiltonians quadratic in the
field momenta (e.g., scalar, electromagnetic and gravitational fields). The
basic underlaying concept, the addition of nonclassical\ momentum
fluctuations to a classical ensemble, carries through from particles to
fields, although significant technical generalizations are needed (see \cite%
{HKR03} for details). In the rest of the paper, I discuss the application of
the exact uncertainty approach to the gravitational field. However, before
applying the exact uncertainty approach to the gravitational field, it will
be necessary to introduce the Hamilton-Jacobi formulation of general
relativity and to define a theory of classical ensembles consistent with
this Hamilton-Jacobi formulation.

A Hamilton-Jacobi formulation for the gravitational field can be defined in
terms of the functional equations 
\begin{equation}
\mathcal{H}=\frac{1}{2}G_{ijkl}\frac{\delta S}{\delta h_{ij}}\frac{\delta S}{%
\delta h_{kl}}-\sqrt{h}R=0,  \label{Eq05}
\end{equation}%
\begin{equation}
\mathcal{H}_{i}=-2D_{j}\left( h_{ik}\frac{\delta S}{\delta h_{kj}}\right) =0.
\label{Eq06}
\end{equation}%
where $R$\ is the curvature scalar and $D_{j}$\ the covariant derivative on
a three-dimensional spatial hypersurface with (positive definite) metric $%
h_{kl}$, and 
\[
G_{ijkl}=\frac{1}{\sqrt{h}}\left(
h_{ik}h_{jl}+h_{il}h_{jk}-h_{ij}h_{kl}\right) . 
\]%
is the DeWitt supermetric (units are chosen so that Newton's gravitational
constant $G=1/16\pi $) \cite{MTW73}.

As a consequence of the Hamiltonian constraint $\mathcal{H}=0$, eq. (\ref%
{Eq05}), and the momentum constraints $\mathcal{H}_{i}=0$, eq. (\ref{Eq06}), 
$S$ must satisfy an infinite set of constraints, numbering four per
three-dimensional point.\ $S$ also satisfies the condition $\frac{\partial S%
}{\partial t}=0$ \cite{B66}. The momentum constraints are equivalent to
requiring the invariance of the Hamilton-Jacobi functional $S$ under spatial
coordinate transformations.\ One may therefore formulate the theory in an
equivalent way by keeping the Hamiltonian constraint, ignoring the momentum
constraints, and requiring instead that $S$ be invariant under the gauge
group of spatial coordinate transformations.

To define classical ensembles for gravitational fields, it is necessary to
introduce some additional mathematical structure: a measure $Dh$ over the
space of metrics $h_{kl}$ and a probability functional $P\left[ h_{kl}\right]
$. A standard way of defining the measure \cite{DW79,HW99} is to introduce
an invariant norm for metric fluctuations that depends on a parameter $%
\omega $,%
\[
\left\Vert \delta h\right\Vert ^{2}=\int d^{n}x\left[ h\left( x\right) %
\right] ^{\omega /2}H^{ijkl}\left[ h(x);\omega \right] \delta h_{ij}\delta
h_{kl} 
\]%
where $n$ is the number of dimensions and 
\[
H^{ijkl}=\frac{1}{2}\left[ h\left( x\right) \right] ^{\left( 1-\omega
\right) /2}\left[ h^{ik}h^{jl}+h^{il}h^{jk}+\lambda h^{ij}h^{kl}\right] 
\]%
is a generalization of the inverse of the DeWitt supermetric (in \cite{DW79}
the particular case $\omega =0$ was considered).\ This norm induces a local
measure for the functional integration given by 
\[
\int d\mu \left[ h\right] =\int \prod_{x}\left[ \det H\left( h\left(
x\right) \right) \right] ^{1/2}\prod_{i\geq j}dh_{ij}\left( x\right) 
\]%
where 
\[
\det H\left( h\left( x\right) \right) \propto \left( 1+\frac{1}{2}\lambda
n\right) \left[ h\left( x\right) \right] ^{\sigma }, 
\]%
and $\sigma =\left( n+1\right) \left[ \left( 1-\omega \right) n-4\right] /4$%
\ (one needs to impose the condition $\lambda \neq -2/n$, otherwise the
measure vanishes). Therefore, up to an irrelevant multiplicative constant,
the measure takes the form 
\begin{equation}
\int d\mu \left[ h\right] =\int \prod_{x}\left[ \sqrt{h\left( x\right) }%
\right] ^{\sigma }\prod_{i\geq j}dh_{ij}\left( x\right) .  \label{Eq07}
\end{equation}%
Without loss of generality, one may set $Dh$ equal to $d\mu \left[ h\right] $
with$\ \sigma =0$, since a term of the form $\left[ \sqrt{h\left( x\right) }%
\right] ^{\sigma }$ may be absorbed into the definition of $P\left[ h_{kl}%
\right] $.

It is natural to require also that $\int DhP$ be invariant under the gauge
group of spatial coordinate transformations. Since the family of measures
defined by eq. (\ref{Eq07}) is invariant under spatial coordinate
transformations \cite{HW99,FP74}, the invariance of $\int DhP$ leads to a
condition on $P$ that is similar to the one required of $S$. To show this,
consider an infinitesimal change of coordinates $x^{\prime k}=x^{k}+\epsilon
^{k}\left( x\right) $ and the corresponding transformation of the metric, $%
h_{kl}\rightarrow h_{kl}-\left( D_{k}\epsilon _{l}+D_{l}\epsilon _{k}\right) 
$. The variation of $\int DhP$ can be expressed as 
\[
\delta _{\epsilon }\int DhP=\int Dh\left[ D_{k}\left( \frac{\delta P}{\delta
h_{kl}}\right) \epsilon _{l}+D_{l}\left( \frac{\delta P}{\delta h_{kl}}%
\right) \epsilon _{k}\right] . 
\]%
Therefore, $\delta _{\epsilon }\int DhP=0$ requires%
\begin{equation}
D_{k}\left( \frac{\delta P}{\delta h_{kl}}\right) =0.  \label{Eq08}
\end{equation}%
or the gauge invariance of $P$. In addition to eq. (\ref{Eq08}), it will be
assumed that $\frac{\partial P}{\partial t}=0$ also holds.

Finally, it should be pointed out that one must factor out the infinite
diffeomorphism gauge group volume out of the measure to calculate finite
averages using the measure $Dh$ and probability functional $P$. This can be
achieved using the geometric approach described in \cite{M95}. This issue
will not be discussed further here, since it does not affect the derivation
of the equations of motion.

An appropriate ensemble Hamiltonian for the gravitational field is given by 
\begin{equation}
\tilde{H}_{c}=\sum_{x}\int DhP\mathcal{H}\sim \int d^{3}x\int DhP\mathcal{H}.
\label{Eq09}
\end{equation}%
The equations of motion derived from eq. (\ref{Eq09})\ are of the form%
\[
\frac{\partial P}{\partial t}=\frac{\Delta \tilde{H}_{c}}{\Delta S},\quad 
\frac{\partial S}{\partial t}=-\frac{\Delta \tilde{H}_{c}}{\Delta P} 
\]%
where $\Delta /\Delta F$ denotes the variational derivative with respect to
the functional $F$. With $\frac{\partial S}{\partial t}=$ $\frac{\partial P}{%
\partial t}=0$, the equations of motion take the form 
\begin{equation}
\mathcal{H}=0,  \label{Eq10}
\end{equation}%
\begin{equation}
\int d^{3}x\frac{\delta }{\delta h_{ij}}\left( PG_{ijkl}\frac{\delta S}{%
\delta h_{kl}}\right) =0  \label{Eq11}
\end{equation}%
Eq. (\ref{Eq10}) is the Hamiltonian constraint, and eq. (\ref{Eq11}) may be
interpreted as a continuity equation.

It is of interest that the interpretation of eq. (\ref{Eq11}) as a
continuity equation leads to a rate equation that relates $\frac{\partial
h_{kl}}{\partial t}$ and $\frac{\delta S}{\delta h_{kl}}$. This follows from
the observation that such an interpretation is only possible if the
\textquotedblleft field velocity\textquotedblright\ $\frac{\partial h_{kl}}{%
\partial t}$\ is related to $G_{ijkl}\frac{\delta S}{\delta h_{kl}}$ in a
linear fashion. Indeed, the most general rate equation for the metric $%
h_{ij} $ that is consistent with the interpretation of eq. (\ref{Eq11}) as a
continuity equation is of the form 
\[
\delta h_{ij}=\left( \gamma G_{ijkl}\frac{\delta S}{\delta h_{kl}}+\delta
_{\epsilon }h_{ij}\right) \delta t 
\]%
where $\gamma $ is an arbitrary function of $x$ (I have include a term $%
\delta _{\epsilon }h_{kl}=-\left( D_{k}\epsilon _{l}+D_{l}\epsilon
_{k}\right) $ which allows for gauge transformations of $h_{kl}$, which is
permitted because the gauge transformations are assumed to leave $\int DhP$
invariant, as discussed before). \ Therefore, one may write the rate
equation for $h_{kl}$ in the standard form 
\begin{equation}
\frac{\partial h_{ij}}{\partial t}=NG_{ijkl}\frac{\delta S}{\delta h_{kl}}%
+D_{i}N_{j}+D_{j}N_{i}.  \label{Eq13}
\end{equation}%
Eq. (\ref{Eq13}) agrees with the equations derived from the ADM canonical
formalism, provided $N$ is identified with the lapse function and $N_{k}$
with the shift vector \cite{W84}.


\section{quantum equations for gravitational fields}

The derivation of the quantum equations of motion from the exact uncertainty
principle follows essentially the same steps as before \cite{HKR03}. It is
first assumed that the field momenta is modified by a stochastic term, so
that. 
\[
\pi ^{kl}=\frac{\delta S}{\delta h_{kl}}+f^{kl} 
\]%
where $f^{kl}$ vanishes on average for all configurations, $\overline{f^{kl}}%
=0$. The classical ensemble Hamiltonian $\tilde{H}_{c}$ is then replaced by
a modified ensemble Hamiltonian $\tilde{H}_{q}$, 
\[
\tilde{H}_{q}=\tilde{H}_{c}+\frac{1}{2}\int d^{3}x\int DhPG_{ijkl}\overline{%
f^{ij}f^{kl}} 
\]%
The form of the modified ensemble Hamiltonian is determined using the same
principles as before (action principle, independence, invariance and exact
uncertainty), leading to the result%
\begin{equation}
\tilde{H}_{q}=\tilde{H}_{c}+\frac{C}{2}\int d^{3}x\int Dh\frac{1}{P}G_{ijkl}%
\frac{\delta P}{\delta h_{ij}}\frac{\delta P}{\delta h_{kl}}  \label{Eq14}
\end{equation}%
where $C$ is a positive universal constant. One can derive an exact
uncertainty relation, defined in terms of generalized covariance matrices
and Fisher information matrices, which connects the statistics of the metric
field and its conjugate momentum fields and implies in turn a
Heisenberg-type inequality (see \cite{HKR03} for derivations of these
results).

If one now defines 
\[
\hbar :=2\sqrt{C},\quad \Psi \left[ h_{kl}\right] :=\sqrt{P}e^{iS/\hbar }. 
\]%
and makes use of the conditions $\frac{\partial S}{\partial t}=$ $\frac{%
\partial P}{\partial t}=0$, it can be shown that\ the Hamiltonian equations
that follow from eq. (\ref{Eq14}) lead to the Wheeler-DeWitt equation%
\[
\left[ -\frac{\hbar ^{2}}{2}\frac{\delta }{\delta h_{ij}}G_{ijkl}\frac{%
\delta }{\delta h_{kl}}-\sqrt{h}R\right] \Psi =0. 
\]

Notice that the exact uncertainty approach specifies a particular operator
ordering for the Wheeler-DeWitt equation. Furthermore, the Wheeler-DeWitt
equation has been derived here using a quantization procedure based directly
on the classical Hamilton-Jacobi formulation, instead of going through the
usual canonical quantization procedure. Therefore, the difficult issues
associated with the requirement of \textquotedblleft Dirac
consistency\textquotedblright\ (i.e., the need to find a choice of operator
ordering \textit{and} regularization scheme that will permit mapping the
classical Poisson bracket algebra of constraints to an algebra of operators
within the context of the Dirac quantization of canonical gravity) have been
avoided. The only ambiguity that remains arises from the need to introduce
some sort of regularization scheme to remove divergences arising from the
product of two functional derivatives acting at the same point.


\section{space-time}

Both the Hamilton-Jacobi theory of the classical gravitational field and the
Wheeler-DeWitt equation involve functional equations defined on a three
dimensional spatial hypersurface, with the group of spatial coordinate
transformations as gauge group. The question then arises of whether it is
possible to relate such formulations to field theories that give a
description of \textit{space-time}.

This is indeed the case for the classical theory, as is well known: one can
show that the Hamilton-Jacobi formulation is equivalent to the usual
formulation in space-time based on the Einstein equations \cite{G69,MTW73}.
The question becomes much more difficult when it comes to the quantized
theory.

In section 4, it was shown that the continuity equation derived from the
classical ensemble Hamiltonian, eq. (\ref{Eq11}), implies the rate equation
for the metric field, eq. (\ref{Eq13}). The same continuity equation and
corresponding rate equation are derived from the quantum ensemble
Hamiltonian, eq. (\ref{Eq14}). This has important implications in the
quantum case, since it suggests a way of evolving away from the spatial
surface using stochastic equations. Although the quantum ensemble
Hamiltonian leads to a rate equation that is identical in form to the
classical rate equation, the interpretation of this equation is somewhat
different, because in the quantum case $\frac{\delta S}{\delta h_{kl}}$
represents an \textit{average} over the non-classical fluctuations of the
momentum. This suggests replacing eq. (\ref{Eq13}) by a \textit{stochastic}
rate equation 
\[
\frac{\partial h_{ij}}{\partial t}\rightarrow NG_{ijkl}\left( \frac{\delta S%
}{\delta h_{kl}}+f^{kl}\right) +D_{i}N_{j}+D_{j}N_{i} 
\]%
and considering the time-evolved spatial metric as being subject to
fluctuations. Furthermore, since the field momenta are subject to
fluctuations, the extrinsic curvature tensor $K_{ij}$, defined by 
\[
K_{ij}=\frac{1}{2}G_{ijkl}\frac{\delta S}{\delta h_{kl}} 
\]%
will also be subject to fluctuations, and should be replaced by a relation
of the form 
\[
K_{ij}\rightarrow \frac{1}{2}G_{ijkl}\left( \frac{\delta S}{\delta h_{kl}}%
+f^{kl}\right) . 
\]%
Since the extrinsic curvature tensor determines how spatial hypersurfaces
fit together, it seems reasonable to expect the emergence of both well
defined evolution equations and a well defined space-time limit after
averaging over the non-classical fluctuations, provided the fluctuations are
not too violent.


\section{discussion}

The approach to quantization of the gravitational field presented here may
be described as a Hamilton-Jacobi quantization of gravity. It differs from
previous approaches that take the classical Hamilton-Jacobi equation as
their starting point in that it incorporates some new elements, in
particular the use of the formalism of ensembles in configuration space and
the postulate of the uncertainty relation as \textit{the} fundamental
element that is needed for the transition from the classical to the quantum
theory. The additional mathematical structure required is minimal. No 
\textit{a priori} assumptions regarding the existence of a Hilbert space
structure, linear operators, wavefunctions, etc. or the use of path
integrals are needed.

The use of ensemble Hamiltonians leads to a unified approach, one in which
both classical and quantum equations are derived using the same mathematical
formalism. In particular, it has been shown that the difference between
classical and quantum ensembles is essentially a `matter of form', being
characterized by the additional term that is added to the classical ensemble
Hamiltonian to quantize the system (i.e., eqs. (\ref{Eq04}) and (\ref{Eq14}%
)). This is of particular interest in the case of gravity, since a unified
approach ensures that the quantum theory has a well defined classical limit.
Furthermore, for any classical quantity that may be expressed as an average
of a functional of $h_{kl}$ and $\frac{\delta S}{\delta h_{kl}}$, one may
define a quantum counterpart by adding nonclassical fluctuations to the
momentum; if this quantity has only linear and quadratic terms in $\frac{%
\delta S}{\delta h_{kl}}$, it will be possible to compute quantum
corrections to the classical quantities in a straightforward way by
averaging over the fluctuations. In this way, classical observables may be
carried over into the quantum theory.

It is remarkable that already at the classical level, the theory of
ensembles in configuration space provides new insight into general
relativity. In particular, as pointed out in section 4, the rate equation
for the metric, eq. (\ref{Eq13}), appears as a direct consequence of the
classical ensemble formalism of gravitational fields. In this way, it takes
only a few steps to show that all 10 Einstein equations follow from the
Hamilton-Jacobi formalism together with the assumption of a well defined
ensemble of gravitational fields. On the other hand, the derivation from the
Hamilton-Jacobi formalism alone \cite{G69}, requires a much more lengthy
derivation. The theory of classical ensembles in configuration space, which
was introduced here as a starting point for the quantization of gravity, is
interesting in its own right, and should be of useful for problems in which
ensembles of gravitational fields need to be taken into consideration.


\section{acknowledgment}

I am grateful to Michael J. W. Hall for his encouragement and many valuable
discussions.

\bibliographystyle{plain}
\bibliography{apssamp}

\end{document}